\documentclass[twocolumn]{aa}
\usepackage{graphicx}
\usepackage{txfonts}
\begin{document}
   \title{Cosmic Shear from STIS pure parallels \thanks{Based on observations made with the NASA/ESA Hubble Space Telescope, 
obtained from the data archive at the Space Telescope Science Institute. STScI is operated by the Association of 
Universities for Research in Astronomy, Inc. under NASA contract NAS 5-26555.}}

   \subtitle{III. Analysis of Cycle 9 pure parallels}

   \author{J.-M. Miralles
          \inst{1}
          \and
          T. Erben\inst{1}
	  \and
	  H. H\"ammerle\inst{1,2}
	  \and
	  P. Schneider\inst{1,2}
	  \and
	  W. Freudling\inst{3}
	  \and
	  N. Pirzkal\inst{3}
	  \and
	  R.A.E. Fosbury\inst{3}
          }

   \offprints{J.-M. Miralles \email{miralles@astro.uni-bonn.de}}

   \institute{Institut f\"ur Astrophysik und Extraterrestrische Forschung der
	Universit\"at Bonn, Auf dem H\"ugel 71, 53121 Bonn, Germany \\   
         \and
             Max-Planck Institute f\"ur Astrophysik, Karl-Schwarzschild Str. 1,
	85748 Garching bei M\"unchen, Germany\\
	 \and
	     ST-ECF, Karl-Schwarzschild Str. 2, 85748 Garching bei M\"unchen, Germany
             }

   \date{}

   \abstract{ 
Following the detection of a cosmic shear signal at the 30$''$ scale using archival parallel data from the STIS CCD camera
onboard HST in H\"ammerle et al. (2002), we analyzed a larger data set obtained from an HST GO pure parallel program. Although this
data set is considerably larger than the one analyzed previously, we do not obtain a significant detection of the cosmic shear 
signal. The potential causes of this null result are the multiple systematics that plague the STIS CCD data, and in particular
the degradation of the CCD charge transfer efficiency after 4 years in space.
   \keywords{cosmology: theory, dark matter, gravitational lensing, 
   large-scale structure of the Universe}
   }

   \maketitle

\def\ave#1{\left\langle #1 \right\rangle}
\def\e{\mathrm{e}}
\def\rh{r_\mathrm{h}} \def\abs#1{\left| #1 \right|}
\def\arcsecf{\hbox{$.\!\!^{\prime\prime}$}}
\def\arcminf{\hbox{$.\!\!^{\prime}$}}
\def\Pg{P^\gamma}
\def\cs{\ave{\overline{\gamma}^2} }
\def\csn{\overline{\gamma}_n^2 }
\def\sigcs{\sigma_{\ave{\overline{\gamma}^2}} }

%

\section{Introduction}

This is the third of a series of articles describing the use of parallel
observations with the STIS CCD camera onboard the HST for the detection of
cosmic shear on scales below one arcmin. 
The first 2 papers (Pirzkal et al. 2001, hereafter PCE01, and H\"ammerle et al. 2002, hereafter HMS02) 
were centered on the analysis of data carefully selected from the HST archive, 
spanning a period between 1997 and 1998. These first 2 papers established that 
the STIS CCD camera used in the CLEAR mode is a useful instrument to measure the value 
of the cosmic shear at scales (30$''$) where ground-based observations are not efficient. The significance of value 
obtained in HMS02 for the rms shear, $\sqrt{\cs}=3.87^{+1.29}_{-2.04}\%$, was limited by the number of 
galaxies in usable fields. To 
strengthen the constraint and decrease the error bars in the shear estimate, we needed more usable fields.
This led us to propose for further observations in parallel mode. This paper concentrates on the analysis of 
data obtained from a Cycle 9 dedicated pure parallel GO proposal. 
Throughout this paper, we employ the same formalism developed in HMS02, and we refer the reader to that paper for 
the complete mathematical description of the terms used. 
The paper is organized as follows: In Sect. 2, we describe the characteristics of the data obtained. Sect. 3 addresses
the field selection and catalog production. Sect. 4 is dedicated to the number counts and sizes of galaxies. In Sect. 5 
we analyse the PSF anisotropy. Sect. 6 describes the shear analysis, including the PSF corrections applied. In 
Sect. 7 we discuss the results obtained, and we concentrate on the understanding of the different effects which influence it.
Finally, in Sect. 8, we summarize our results and try to offer a perspective on the future works using STIS and ACS.


\section{Data}


   \begin{figure*}
   \centering
   \includegraphics[width=8cm]{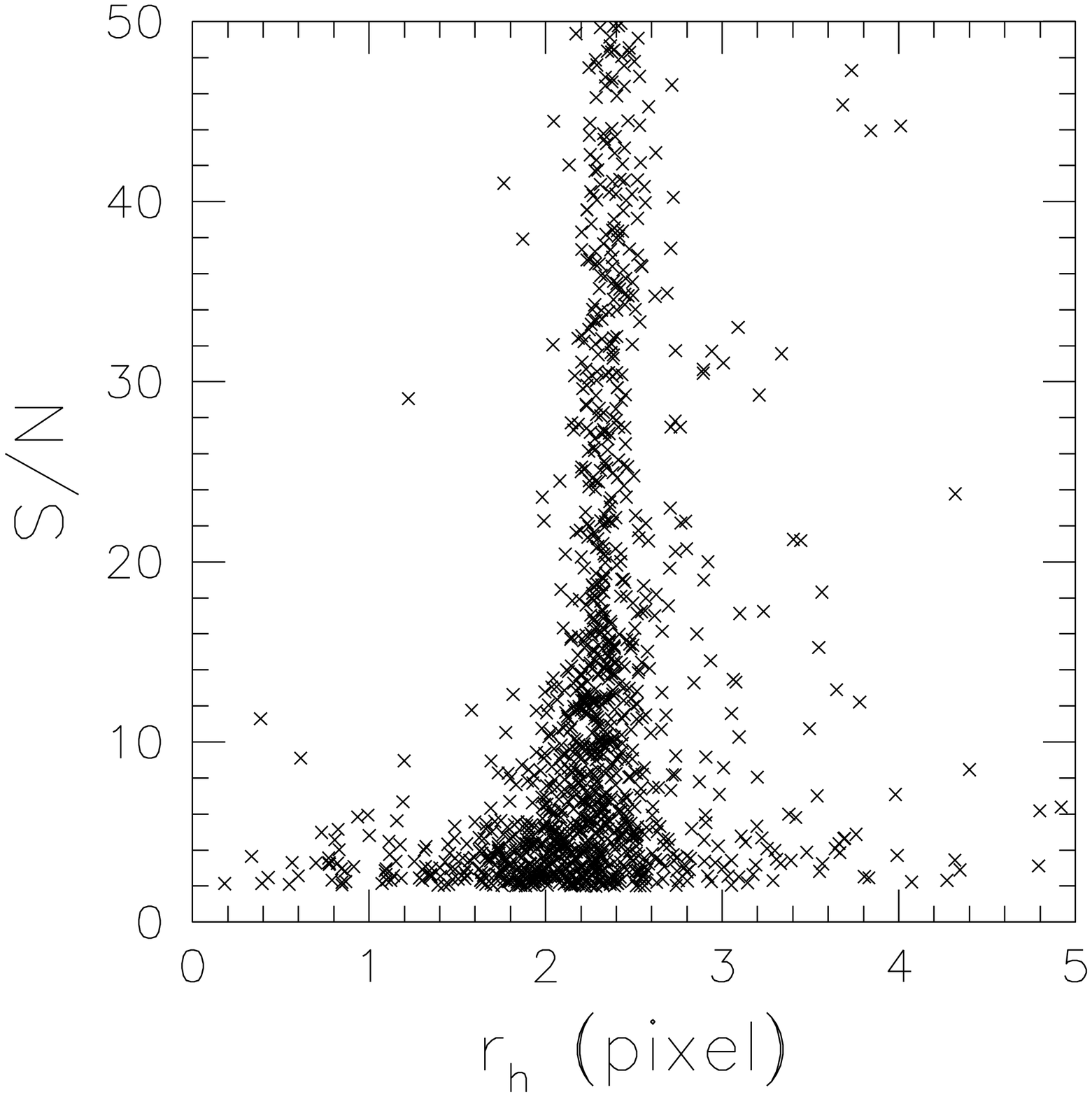}
   \includegraphics[width=8cm]{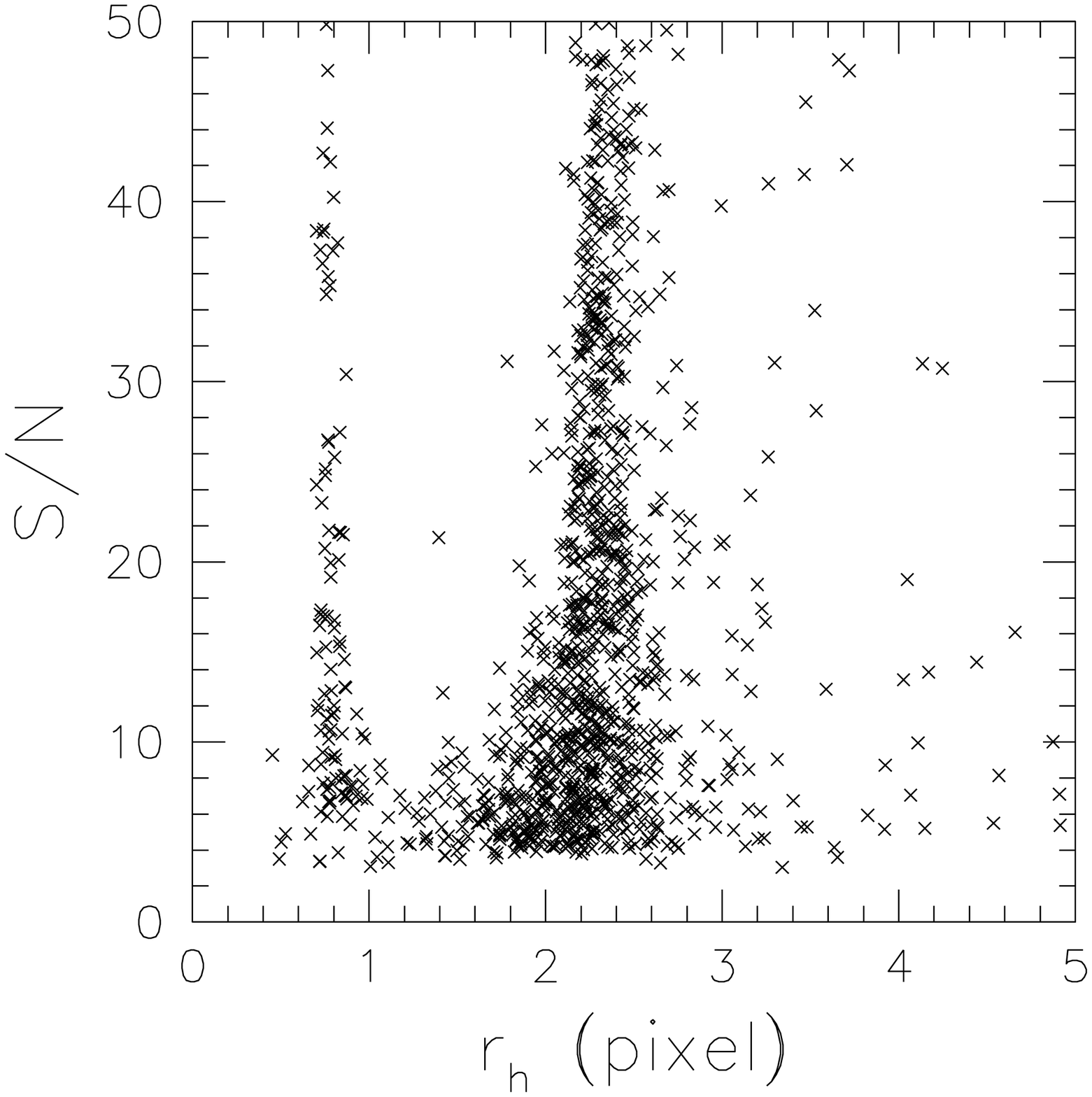}
 \caption{Half-light radius in STIS subsampled pixels vs. S/N for 2 star fields. On the left panel the 
association had a dither pattern between members, while on the right the offsets between members were 
less than 1 pixel. The strip at $r_\mathrm{h} < 1$ on the right panel indicates the presence of individual hot pixels. The
strip at 2.1$< $ $r_\mathrm{h} <$ 2.6 is populated by stars. }
    \end{figure*}

The data which was used for the analysis in this paper was obtained as part of parallel GO proposal 
(8562+9248 P.I.: P. Schneider) from Sept. 24th 2000 until May 16th 2001. 
The latter date corresponds to the moment when a fuse in the primary power
feed of STIS blew up, stopping instrument operations until July 2001 when STIS was 
repowered through the secondary power line.

A total of 3511 individual datasets were obtained during this period.  Each individual image 
was bias substracted and flat-fielded automatically on request by the data archive at the Space Telescope 
Science Institute using the On-the-Fly Calibration process (\cite{crabtree}) with the best available calibration data. 
The images were then associated following the definition given in PCE01 
into 575 STIS associations (\cite{micol}). Each association includes data which were taken consecutively 
during a single telescope visit using the same telescope roll angle, and which were 
offset by no more than one quarter of the field-of-view. The relative offsets between members of the 
associations were computed using the available jitter data and refined using the same iterative 
cross-correlation technique as described in PCE01. 
Each image in the association was drizzled using the IRAF STSDAS DRIZZLE procedure in a subsampled grid with a scale of 0.5 
input pixels and offset to register the different members.
The final associated image was obtained by the median average of the members of the association using the
IRAF IMCOMBINE procedure. During this step, we used CRREJECT to reject pixels with a deviation
larger than 3$\sigma$ from the mean value. This last step was necessary for this dataset since it
was no longer taken in CR-SPLIT mode as it was the case for the data used in PCE01 and HMS02. 

Of these 575 associations, 568 have a total exposure time larger than 400s. The mean exposure 
time for an association is about 2500s. This has to be compared to the mean exposure time of the 
archival data associations which was about 2000s.
Another intrinsic difference between the archival data analyzed and the newly obtained data is 
that it was no longer obtained in the CR-SPLIT mode and therefore we need at least 2 consecutive 
single exposures of the same field to be able to properly remove cosmic rays. We limited our analysis 
then to the 484 associations for which we have at least 2 members. All those datasets can be found
on http://www.stecf.org/projects/shear/

Still we can see from Fig. 1 that some images, for which the members are not dithered by more than 1 pixel, 
are contaminated by a number of hot pixels. Between 1997-1998, when the archival data was taken, and the
period 2000-2001, when the cycle 9 data was obtained, the number of hot pixels was increased by a factor of 4 (see 
Fig. 1 in Proffitt et al. 2002a). This increase in number has made it very difficult for the hot pixels
to be cleaned up from non-dithered images. 
Those hot pixels which are isolated can be identified as they form a strip with half-light radius of less 
than one subsampled STIS pixel.
Therefore they can be rejected when we select objects (stars and galaxies) by their half-light radius. However, a number
of hot pixels will lay inside objects, affecting their shape measurement. Because the distribution of those 
pixels is random, when averaging, they will
affect the noise in our measurement, increasing the dispersion of ellipticities.  

Besides this higher contamination from hot pixels, the degradation in the charge transfer efficiency 
(CTE) between 1997 and 2000 has to be considered also. The radiation to which the CCD is subjected in space degrades
the capability of the charges to be successfully moved through adjacent pixels. This effect is characterized
by a loss of flux in objects with increasing distance to the read-out amplifier, which for our data is 
located at the top-right. This loss of efficiency is particularly strong in the Y direction as shown by
Goudfrooij et al. (2002) and negligible in the X direction. This leads to the apparition of visible trails 
parallel to the Y direction for bright stars as can be seen in Fig. 2. The effect of the degradation of the CTE is
discussed in more detail in Sect. 7.3. 

   \begin{figure}
   \centering
   \includegraphics[width=8.5cm]{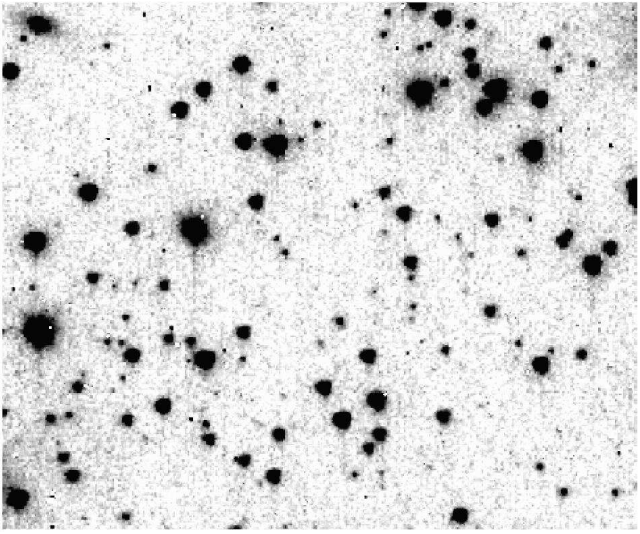}
 \caption{Snapshot of one of the starfields where the CTE degration can be clearly seen in the form of trails 
originating from the stars and parallel to the Y direction.}
    \end{figure}

\section{Field selection and catalogue production}

The co-added associations were inspected visually and classified as either star fields or galaxy
fields in the same fashion as done in PCE01. The final selection included 210 galaxy fields and 110 star fields. 
The complete list of star and galaxy fields can be found on http://www.stecf.org/projects/shear/

The fields were analyzed in the same way as in HMS02 which is summarized in the following:
SExtractor (Bertin \& Arnouts 1996) catalogues were produced using the parameter file which can be found at 
http://www.stecf.org/projects/shear/sextractor. We discarded all objects flagged internally by SExtractor due to 
problematic deblending and/or thresholding, and we removed all objects located at less than 25 subsampled STIS pixels from
the edges of the images which are highly noisy. Furthermore, we applied manual masks to problematic regions like
diffraction spikes from saturated stars. 
IMCAT (Kaiser et al. 1995, Hoekstra et al. 1998, Erben et al. 2001) catalogues were produced parallely and merged 
with the cleaned SExtractor catalogues, requiring an object to be uniquely detected in both catalogues within a radius 
of 125 mas (5 subsampled STIS pixels) from the coordinates as defined by the SExtractor catalog. 
In the final merged catalog, we use the size (half-light radius $r_\mathrm{h}$) and shape parameters (Luppino \& Kaiser 1997; 
Hoekstra et al. 1998) from IMCAT, with the position and the magnitude defined by SExtractor.

We identified stars with the same parameters as in the data presented in HMS02: 2.1$<$ $r_\mathrm{h}<$2.6 and S/N$>$10.
For the galaxies, we applied a slightly more conservative criterion of $r_\mathrm{h}>$2.7 and S/N$>$5.

\section{Galaxy number counts and sizes}

   \begin{figure}
   \centering
   \includegraphics[width=8.5cm]{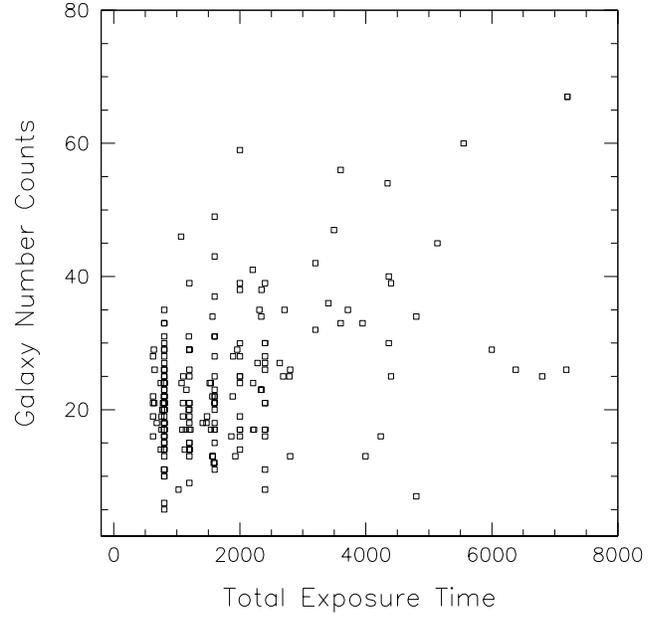}
   \caption{Number of detected galaxies per STIS association as a function of total exposure time in seconds
for the 210 galaxy fields.}
    \label{galexp}
    \end{figure}

   \begin{figure}
   \centering
   \includegraphics[width=8.5cm]{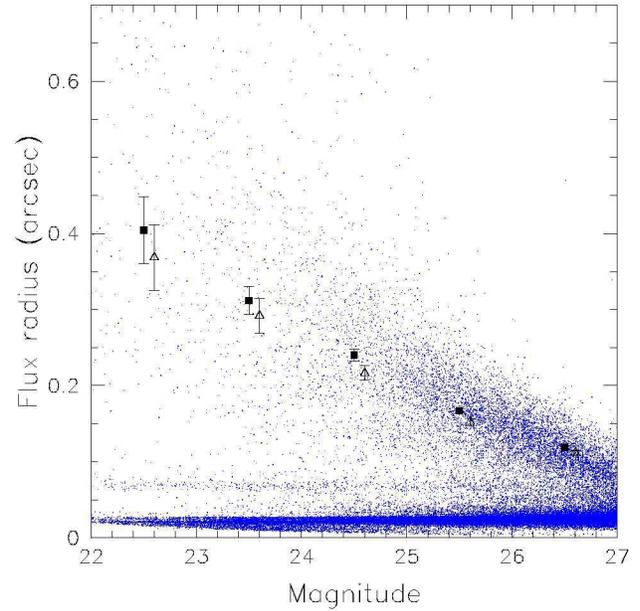}
 \caption{Half-light radius as measured by SExtractor for the objects detected in the galaxy fields as a function of the
CLEAR filter AB magnitude. The horizontal strip at a half-light radius between 0\arcsecf05 and 0\arcsecf08 is caused 
by stars and other unresolved objects. The strip at the half-light radius between 0\arcsecf025 and 0\arcsecf03 is a product of
the single noise pixels, mostly hot-pixels. The average size of galaxies (objects with a half-light 
radius greater than 0\arcsecf08) per magnitude bin are indicated by the full squares. The error bars 
represent the 3$\sigma$ level in the error of the mean. The open triangles represent the same for
the fields studied in PCE01 and HMS02.}
              \label{rhmag}

    \end{figure}
In Fig. 3, we plot the number of detected galaxies per associated image as a function of total 
exposure time. When compared with the archival data, the average total exposure time is higher, 2500s, 
leading to a higher mean number of galaxies per field of 24 (35 gal/arcmin$^2$). The average exposure time is
very close to the optimal integration time deduced from the 
archival data up to which the number of galaxies detected per field rises steadily as a function of integration time. 
After that, the number of galaxies detected rises more slowly because the intrisic flattening
of galaxy number counts at faint magnitudes and because of those faint objects become too small to be resolved by STIS. 

This effect can be seen also in Fig. 4, where the half-light radius measured by SExtractor
of the objects in galaxy fields as a function of the magnitude is plotted. The average half-light radius for galaxies 
varies from 0\arcsecf37 for magnitude bin 22-23 to 0\arcsecf1 for magnitude bin 26-27. Those values are consistent with what
was found in PCE01 and in Gardner \& Satyapal (2000). 
Beyond magnitude 27, most of the objects become unresolved for STIS and therefore cannot be used for our analysis.

\section{Analysis of the PSF anisotropy}

Since the expected distortion of image ellipticities on the STIS angular scale should be a few percent, any
instrumental distortion and other causes of PSF anisotropy need to be controled to an accuracy of better than 1\%.
We showed in HMS02 that the STIS PSF anisotropy remained remarkably stable and was sufficiently small (less than 1\%) in amplitude 
between June 1997 and October 1998.
We analyzed the PSF properties of the newly obtained data to check whether this remained the case. We show in
Fig. 5 the mean ellipticities of stars, $e_1$ and $e_2$ as defined in Eq. (3) of HMS02, over the whole
field as a function of the epoch. The variation of the PSF anisotropy in time is similar to what was found in
HMS02 for the period 1997-1998. For the stars, the mean $e_1$ is about 1\% and the mean $e_2$ is close to 0, 
with a dispersion of about 1\%. Therefore, we consider the PSF anisotropy to be constant over the period 
of time covered.

We investigated also the spatial variation of the PSF within invidual fields. We show a characteristic star field 
in Fig. 6. We applied the same procedure described in HMS02 [Eqs. (5) and (6)] of fitting a second-order
polynomial to the ellipticity of a star at each star position on the CCD. Comparing this with Fig. 4 and 5 of 
HMS02, we find that the anisotropy patterns and their dispersions are similar for both datasets.  

   \begin{figure*}
   \centering
   \includegraphics[width=8cm]{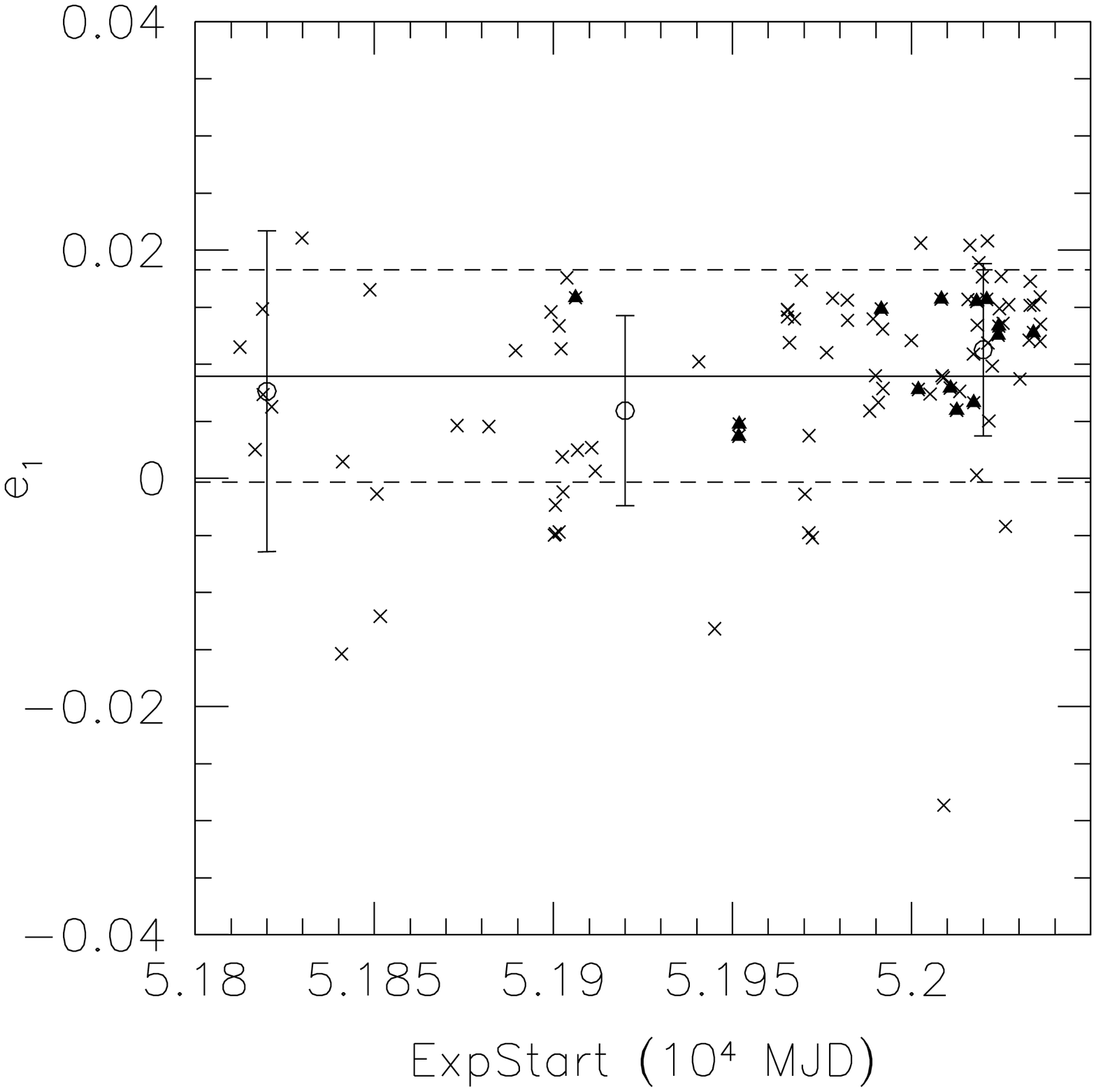}
   \includegraphics[width=8cm]{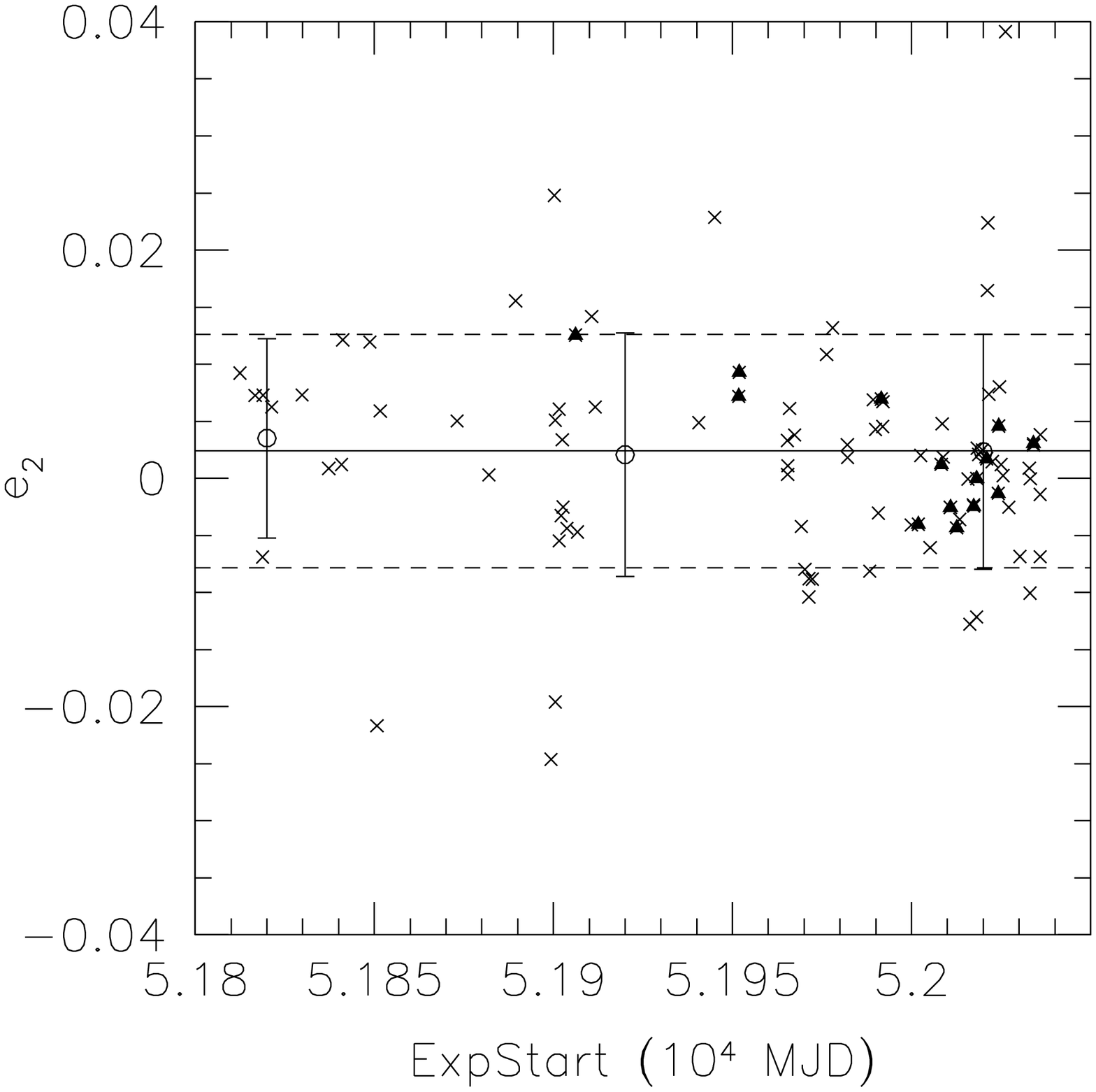}
 \caption{Mean ellipticity components $e_1$, on the left, and $e_2$, on the right, of the star fields
vs. epoch of the observations in Modified Julian Date. The straight solid line shows for the mean over all the 
fields, with the dashed lines showing the 1$\sigma$ dispersion. The circles show the mean for star 
fields in 3 different time bins together with the 1$\sigma$ error bars for those values. The black triangles indicate
the fields that were selected for the PSF correction.}
    \end{figure*}
 
   \begin{figure*}
   \centering
   \includegraphics[width=15cm]{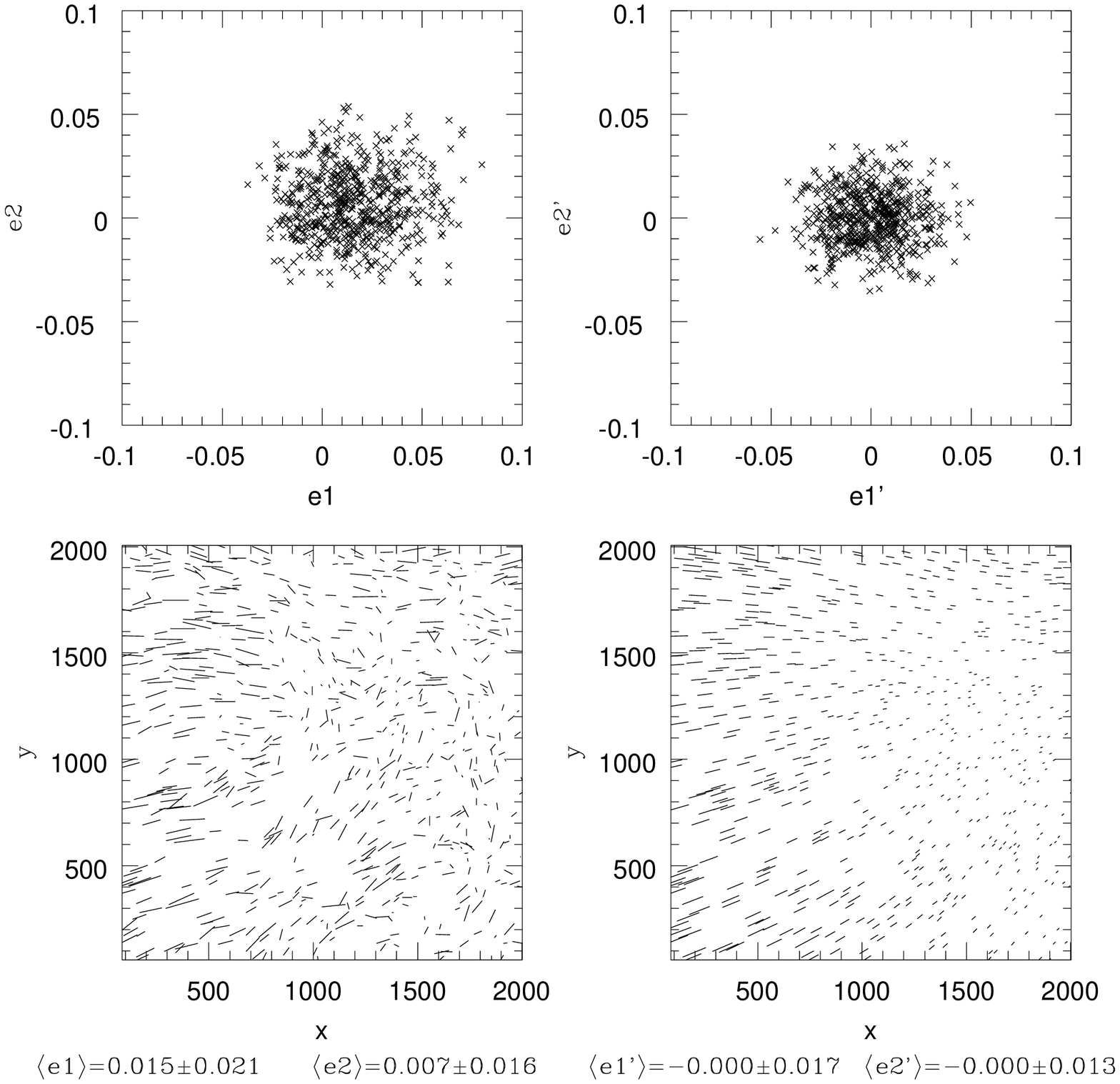}
 \caption{For the star field o6fx9j010\_2.ass, we show the distribution of the ellipticities of stars
before (top left) and after (top right) correcting for the PSF anisotropy using the fitted second order 
polynomial plotted on the bottom right panel. The bottom left panel shows the original spatial distribution of 
the ellipticities. The length of the sticks indicates the modulus of the ellipticity and the orientation
gives the position angle.}
    \end{figure*}


\section{Shear analysis}
\subsection{PSF anisotropy correction}

From the 110 star fields obtained, we selected 14 for the PSF correction of galaxies. This selection
was done with the same criteria as for the archival data: good spatial coverage of stars and small 
intrinsic dispersion in star ellipticities.  This allows us to have a good fit to the anisotropy pattern
and to minimize the noise in the PSF correction. The galaxy ellipticities were corrected for anisotropy; the 
formalism can be found in Eqs. (14), (15), (16) and (17) of HMS02.
Each galaxy in each galaxy field is corrected by one of the selected star fields. Averaging
over all the galaxies, we obtain a mean anisotropy corrected ellipticity for each galaxy field with a particular 
star field correction. Then, averaging over the 14 different star fields corrections, we obtain an average 
value for the mean anisotropy corrected ellipticity for each galaxy field. 

In Fig. 7, the mean ellipticity for the galaxies in all the galaxy fields is shown with and without
the PSF anisotropy correction. The difference between the mean anisotropy corrected and uncorrected ellipticities is less than 1\%. 
One trend emerges though, as the anisotropy correction shifts the mean ellipticities into the negative $e_1$ direction. This is not
unexpected since there is a positive $e_1$ component on the star fields as seen in Fig. 5, which was also
present in the archival data. 

One big difference arises between this dataset and the archival data. The mean ellipticity,
corrected or uncorrected, over all the galaxy fields is no longer compatible with 0 as it
should be if the shear due to the large-scale structure is uncorrelated on the different galaxy fields,
and the galaxies are oriented randomly intrinsically. We find $\ave{e_1} = (-1.03\pm0.40) \%$ and 
$\ave{e_2} = (-0.32\pm0.38) \%$ for the uncorrected ellipticities and $\ave{e^\mathrm{ani}_1} = (-1.29\pm0.40)\%$ 
and $\ave{e^\mathrm{ani}_2} = (-0.40\pm0.38)\%$ for the anisotropy corrected ellipticites, where the 
errors given are calculated as errors on the mean as $\sigma/\sqrt{N}$, not taking into account the 
dispersion from the different PSF corrections. The deviation from zero for the $e_1$ component is 3$\sigma$ significant and 
indicates the presence of a systematic effect in the galaxy fields.

\begin{figure*}
   \centering
   \includegraphics[width=15cm]{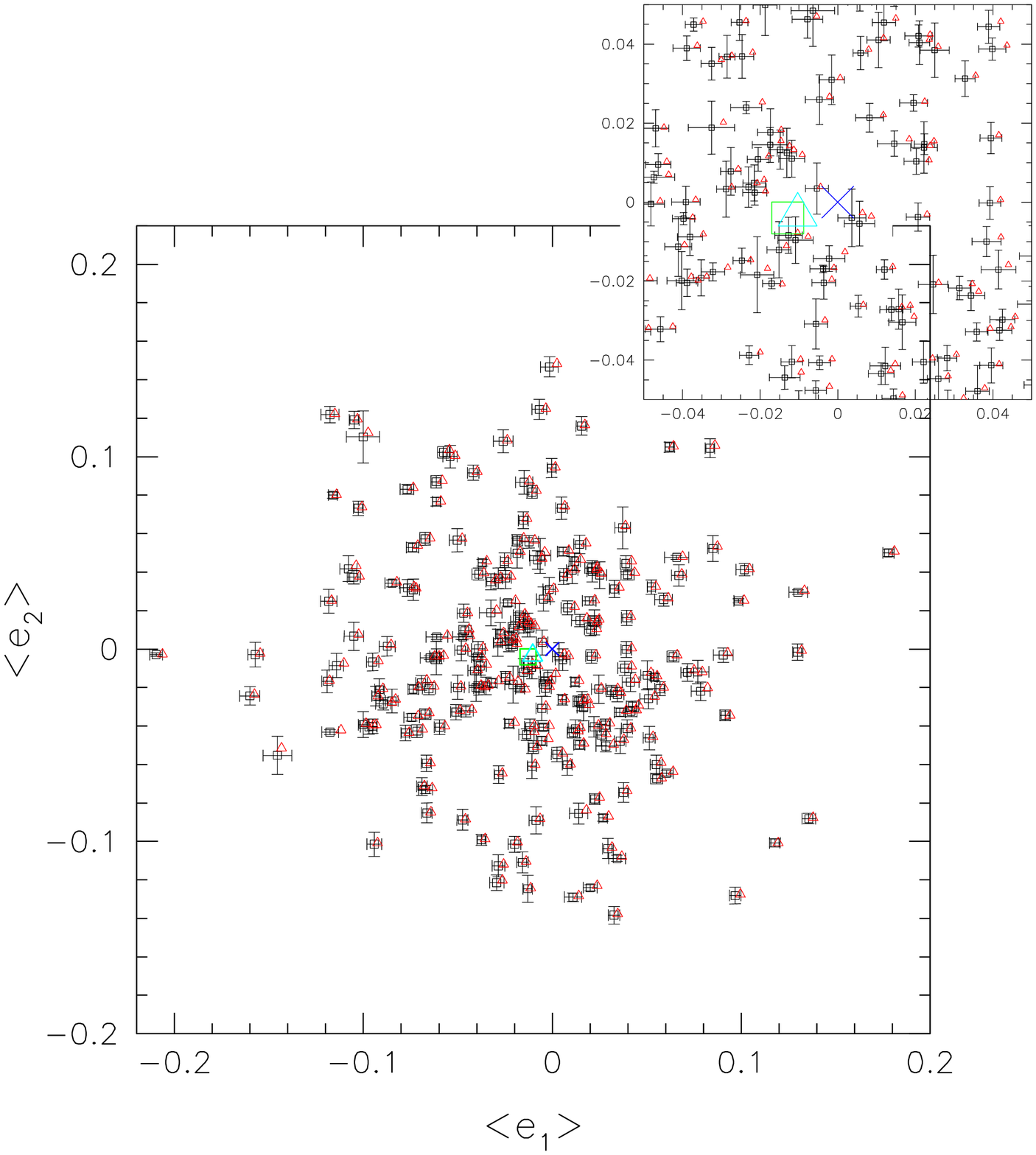}
 \caption{For the 210 galaxy fields, we plot the mean uncorrected ellipticity of galaxies (triangles) as
well as the mean anisotropy corrected ellipticity (squares). The error bars on the squares indicate 3 times
the dispersion of the field-averaged corrected ellipticities with the different PSF model fits. The error
on the mean is in reality smaller than the symbols used. The big triangle and the big square at the center
show the mean over all galaxy fields of respectively the uncorrected and corrected mean ellipticities. Their 
size represent the 1$\sigma$ errors on the mean. The cross marks the 0,0 point. For clarity, a zoom on the 
central part is showed on the upper right corner.}
\end{figure*}

\subsection{Smearing correction and fully corrected ellipticities}
The fully corrected ellipticity is obtained from the anisotropy corrected ellipticities by 
$\e^\mathrm{iso} = (\Pg)^{-1} e^\mathrm{ani}$ [Eq. (18) in HMS02], 
where $\Pg = P_\mathrm{sh} - (P_\mathrm{sh}/P_\mathrm{sm})^\star P_\mathrm{sm}$ [Eq. (15) in HMS02]. We apply a scalar inversion of this 
tensor $(\Pg)^{-1}=2/{\rm{tr}}\Pg$ which is less noisy than the full tensor inversion as demonstrated in Erben et al. (2001).
Therefore to compute the fully corrected ellipticity of each galaxy, we need to calculate $\Pg$ for each galaxy, and for this purpose 
we need to estimate the ratio of the shear and smear susceptibility tensors $(P_\mathrm{sh}/P_\mathrm{sm})^\star$ from the 
light profile of the stars with the filter scale of the galaxy [see Eq. (1) in HMS02].  
We find no spatial variation of $(P_\mathrm{sh}/P_\mathrm{sm})^\star$ for any filter scale over the selected star fields as
it was the case for the archival data. 
But there are variations between the different star fields, closely related to the average size of the stars, 
r$_\mathrm{h}=2.35 \pm 0.25$ subsampled STIS pixels, which are compatible from field to field within the error bar. 
Taking the mean over all the 14 selected star fields for different filter scales, we show in Fig. 8 the dependence of 
$(P_\mathrm{sh}/P_\mathrm{sm})^\star$ on the filter scale. This ratio increases with the filter scale and becomes
 constant for large 
objects.
This is intuitively expected since small objects are going to be more affected by the PSF smearing and therefore
are going to have a larger $P^*_\mathrm{sm}$. The dependence shown for the Cycle 9 data is similar to the one
for the archival data and the values are compatible within the error bars. 

\begin{figure}
   \centering
   \includegraphics[width=8.5cm]{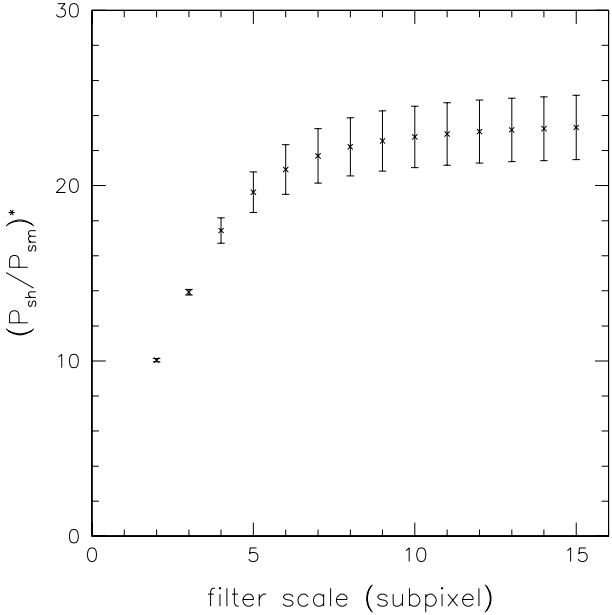}
 \caption{The mean of $(P_\mathrm{sh}/P_\mathrm{sm})^\star$ over all
 14 stars fields which were used for the PSF correction is shown for
 different filter scales. The error bars show the dispersion between
 different star fields.}
\end{figure}

Since to fully correct the ellipticities, one has to divide by $\Pg$, objects with small values of $\Pg$ can 
get unphysically large ellipticities. Those can dominate the cosmic shear signal even after the introduction of 
a weighting scheme which is discussed in the following section. Therefore, we decided, as in HMS02, to introduce a 
cut in $\Pg$, requiring that ``good'' galaxies should have $\Pg > 0.2$ and $\abs{e}<1$. The effects of this cut are discussed in Sect. 7.1.
As for the archival data, each galaxy field is corrected with each of the selected starfields. The dispersion of the corrected galaxy 
ellipticities using the different starfield corrections is only about 1\%. The mean PSF fully corrected ellipticity is then 
obtained by averaging over all the individual corrections. 
When averaging over all the galaxies, we find that $\ave{e^\mathrm{iso}_1} = (-1.41\pm0.41)\%$
and $\ave{e^\mathrm{iso}_2} = (-0.34\pm0.41)\%$ where the error quoted is the error on the mean given by 
the statistical dispersion of the ellipticities divided by the square root of the total number of galaxies. 
The dispersion of ellipticities is found to be 26$\%$ for both components as it was the case for the archival data.

\subsection{Weighting scheme}
The PSF corrections, anisotropy and smearing, applied to the galaxy ellipticities amplify the 
measurement error of these values. This can lead for certain objects to unphysical ellipticities. 
The goal of the weighting is to minimize the impact of those high-ellipticity objects which
are most likely to originate from noise. 
The adopted procedure is the same as we used for the data on HMS02. Since we expect galaxies with small sizes
 and/or low S/N to be the most sensitive to noise, we search for each galaxy the
20 nearest neighbours in the (r$_\mathrm{h}$, S/N) parameter space (Erben et al. 2001) and calculate the dispersion
of their fully-corrected ellipticities which we call $\sigma_\mathrm{NN}$. The weight is then just defined as
$w_\mathrm{NN}=1/{\sigma_\mathrm{NN}}^2$. 
Even in the case of a perfect measurement and a perfect PSF correction, there would be a dispersion in the shear 
estimate, $\sigma_\mathrm{g}$,  because of the contribution of the intrinsic ellipticity dispersion of the galaxy 
population $\sigma_\mathrm{s}$. Therefore, since $\sigma_\mathrm{NN}$ is an estimate of $\sigma_\mathrm{g}$ , it can be never lower 
than $\sigma_\mathrm{s}$.
This leads to an upper limit for the weights given by: 
\begin{equation}
w_\mathrm{NN}^\prime = \mathrm{min} \left( \frac{1}{\sigma_\mathrm{NN}^2},
\frac{1}{\sigma_\mathrm{s}^2} \right)
\end{equation}
with $\sigma_\mathrm{s}$ = $\sqrt{2} \times 26\%$ as measured from the distribution of the corrected galaxy ellipticities.

\subsection{Cosmic Shear estimation}

The fully corrected ellipticity is an unbiased (but noisy) estimate of the shear. We use it 
to calculate an estimator of the shear variance for each of the  $n$ galaxy fields by calcultating a weighted average of the 
galaxy ellipticities $\overline{\gamma}^2_n:= \frac{ \sum_{i\neq j} w_{in} w_{jn} \e_{in} \e_{jn}^*} {\sum_{i\neq j} w_{in} w_{jn}}$ 
[Eq. (9) of HMS02]. 
By calculating the weighted average over all the galaxy fields, we obtain the estimated cosmic shear for 
our data $\cs = \frac{\sum{N_n\overline{\gamma}^2}}{\sum{N_n}}$ [Eq (11) in HMS02], where $N_n$ is the number of galaxies per field. 
The associated statistical error $\sigcs$ is defined in Eq. (21) of HMS02. 
The results are summarized in Table 1 for all selected galaxy fields and for fields with more than 10
or more than 15 galaxies per field, with different cuts in $\Pg$ and with different weightings. 

The cosmic shear estimator for the analyzed data is negative or close to 0 in almost all the cases for
different selections and weighting schemes of the galaxies and the galaxy fields.
For the 121 fields in HMS02, we found $\cs = (14.96 \pm 11.61) \times 10^{-4}$ using all fields and all galaxies with $\Pg > 0.2$,
while we find for the 210 galaxy fields of cycle 9 data and using the same selection and correction criteria 
$\cs = (-7.16 \pm 5.13) \times 10^{-4}$. This value is certainly surprising since it is negative for a value that is positively defined. 
But one has to remember that we compute only an estimator of the real shear variance which even if it is unbiased can be found negative 
if it is dominated by noise and systematics.

The variance of the cosmic shear in a circular aperture $\theta/2$ can also be calculated from the correlation 
function $\ave{\gamma\gamma}$ for the scale $\theta$ (Bartelmann \& Schneider, 2001) in an indepedent way from the estimator 
$\cs$. As shown in Fig. 9, where we plot the integrated values for 
the correlation function, $\ave{\gamma\gamma}$ is consistenly negative confirming the result found from the cosmic shear estimator. 
It has to be noted though that since we plot the integrated correlation function, points are not independent from each other. 

Another way to estimate the significance of the estimator is to randomize the orientation of the galaxies 
in all the galaxy fields and weight them in the same way than for the measured result. The probability distribution function for 
the cosmic shear estimator obtained in that way is shown on Fig. 10. In 93.4$\%$ of the randomizations, the value
is higher than the measured one, confirming then that the negative value for the cosmic shear estimator is not significant 
statistically. The full width at half maximum of the distribution is about $5 \times
10^{-4}$ which agrees with the errors estimated from the intrinsic ellipticity distribution, 
$\sigma_\mathrm{intr}= \frac{\sigma_\mathrm{s}^2} {N \sqrt{N_\mathrm{f}}} = 4 \times 10^{-4}$, 
where $N$ is the average number of galaxies per field plus the error from the cosmic variance 
$\sigma^2_\mathrm{cv} = \frac{\cs}{\sqrt{N_\mathrm{f}}} = 5\times 10^{-5}$ .

\begin{figure}
   \centering
   \includegraphics[width=9cm]{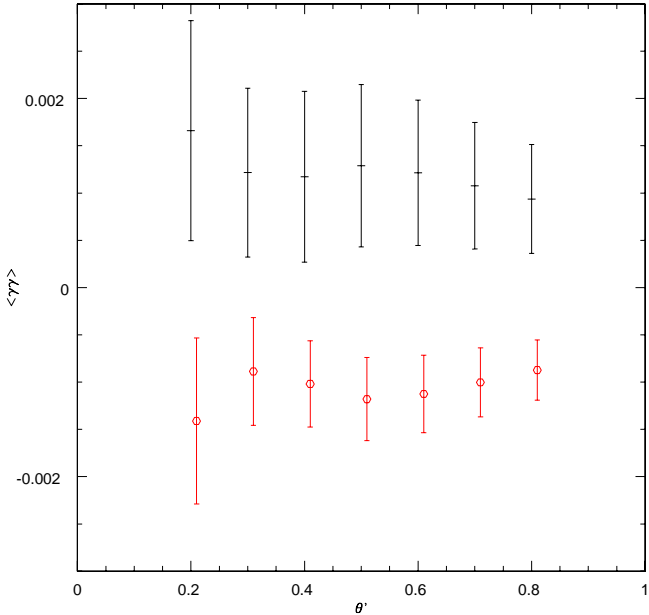}
\caption{Integrated correlation function $\ave{\gamma\gamma}$ as a function
of the scale in arcminutes for the Cycle 9 data (data points with a circle) and
for the archival data of HMS02.}
\end{figure}

\begin{figure}
   \centering
   \includegraphics[width=9cm]{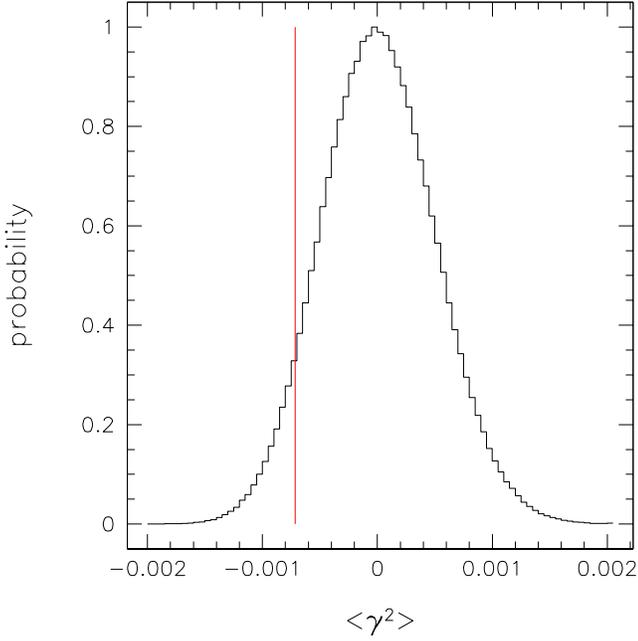}
\caption{Probability distribution of the cosmic shear estimator
 calculated from the Cycle 9 data by randomizing the orientations of galaxies
 used for the cosmic shear analysis in all galaxy  fields. Individual
 galaxies are
 weighted with $w^\prime_\mathrm{NN}$ and the galaxy fields are weighted by
 the number of galaxies on each field. The shape of the distribution
 looks very  similar to the one calculated for the archival data, however, it is much
 narrower than for the archival data which is due to the higher total
 number of  galaxies used in the measurement (note the different
 scalings of the plots).  The vertical line indicates the
 measured cosmic shear value.}
\end{figure}

\begin{table*}
\caption{Results for the cosmic shear estimates and errors for
different  minimum number of galaxies per field.
$N$ is the number of  galaxies per field, $N_\mathrm{f}$ is the number
of galaxy fields with $N\ge N_\mathrm{min}$.
The first block shows the results for different cuts in $\Pg$, where
we weight individual galaxies with $w=w^\prime_\mathrm{NN}$ and the galaxy
fields with $W_\mathrm{f}=N$.
In the next blocks the results are given for weighting individual
galaxies or not, and for applying
different weights to galaxy fields, weighting them equally
($W_\mathrm{f}=1$), weighting by the number of galaxies per field
($W_\mathrm{f}=N$) or by the square of the number of galaxies
($W_\mathrm{f}=N^2$).
The last two blocks show the results for different minimum sizes of the
selected galaxies and when using only fields which were dithered by
more than $1$ pixel.
Note that we repeat the result with $\rh>2.7$, $\Pg>0.2$,
$w=w^\prime_\mathrm{NN}$, $W_\mathrm{f}=N$ in each block (except the
last) for easier comparison of the results.
}
\center
\begin{tabular}{|l|rrr|rrr|rrr|}
  \noalign{\smallskip}
  \noalign{\smallskip}
\hline
 & \multicolumn{3}{c|}{all} & \multicolumn{3}{c|}{$N\ge10$} &
   \multicolumn{3}{c|}{$N\ge15$}\\
 & $N_\mathrm{f}$ & $\cs$ & $\sigcs$ &
   $N_\mathrm{f}$ & $\cs$ & $\sigcs$ &
   $N_\mathrm{f}$ & $\cs$ & $\sigcs$ \\
 && $\times 10^{4}$ & $\times 10^{4}$ && $\times 10^{4}$ & $\times 10^{4}$ &&
    $\times 10^{4}$ & $\times 10^{4}$  \\
\hline
\hline
\multicolumn{10}{|c|}{different cuts in $\Pg$;
$w=w^\prime_\mathrm{NN}$; $W_\mathrm{f}=N$; $\rh>2.7$}\\
\hline
$\Pg>0.0$ & $210$ &$ -7.92$ &$ 5.04$ &$190$ &$ -6.48$ &$ 5.04$ &$142$ &$ -1.33$ &$ 5.05$\\
$\Pg>0.1$ & $210$ &$ -7.19$ &$ 5.03$ &$190$ &$ -5.46$ &$ 5.06$ &$137$ &$  0.61$ &$ 5.17$\\
$\Pg>0.2$ & $210$ &$ -7.16$ &$ 5.13$ &$184$ &$ -6.04$ &$ 4.97$ &$130$ &$  1.44$ &$ 5.36$\\
\hline
\hline
\multicolumn{10}{|c|}{different weighting of individual galaxies;
$\Pg>0.2$; $W_\mathrm{f}=N$; $\rh>2.7$}\\
\hline
$w=1$                    & $210$ &$ -4.95$ &$ 5.51$ &$184$ &$ -3.21$ &$ 5.40$ &$130$ &$  4.11$ &$ 5.76$  \\
$w=w_\mathrm{NN}$        & $210$ &$ -8.72$ &$ 4.96$ &$184$ &$ -8.14$ &$ 4.74$ &$130$ &$ -0.59$ &$ 5.08$  \\
$w=w^\prime_\mathrm{NN}$ & $210$ &$ -7.16$ &$ 5.13$ &$184$ &$ -6.04$ &$ 4.97$ &$130$ &$  1.44$ &$ 5.36$  \\
\hline
\hline
\multicolumn{10}{|c|}{different weighting of galaxy fields; $\Pg>0.2$;
$w=w^\prime_\mathrm{NN}$; $\rh>2.7$}\\
\hline
$W_\mathrm{f}=1$      & $210$ &$-13.77$ &$ 5.31$ &$184$ &$-10.41$ &$ 4.95$ &$130$ &$  1.82$ &$ 5.43$ \\
$W_\mathrm{f}=N$      & $210$ &$ -7.16$ &$ 5.13$ &$184$ &$ -6.04$ &$ 4.97$ &$130$ &$  1.44$ &$ 5.36$ \\
$W_\mathrm{f}=N^2$    & $210$ &$ -3.12$ &$ 6.08$ &$184$ &$ -2.79$ &$ 5.98$ &$130$ &$  1.06$ &$ 6.18$ \\

\hline
\multicolumn{10}{|c|}{different cuts in half-light radius; $\Pg>0.2$;
$W_\mathrm{f}=N$; $w=w^\prime_\mathrm{NN}$}\\
\hline
$\rh>2.7$&$210$ &$ -7.16$ &$ 5.13$ &$184$ &$ -6.04$ &$ 4.97$ &$130$ &$  1.44$ &$ 5.36$  \\
$\rh>3  $&$210$ &$ -7.31$ &$ 6.01$ &$181$ &$ -7.56$ &$ 5.26$ &$120$ &$  1.90$ &$ 5.21$  \\
$\rh>4  $&$210$ &$ -8.41$ &$ 8.02$ &$148$ &$ -3.98$ &$ 7.79$ &$ 78$ &$  3.66$ &$ 8.78$  \\
$\rh>5  $&$210$ &$ -5.20$ &$14.41$ &$ 83$ &$ 12.19$ &$11.91$ &$ 34$ &$ 15.24$ &$19.51$  \\
\hline
\hline
\multicolumn{10}{|c|}{using only fields dithered by $>1$ pixel; $\Pg>0.2$;
$W_\mathrm{f}=N$; $w=w^\prime_\mathrm{NN}$; $\rh>2.7$}\\
\hline
 &$95$  &$-2.39$  &$ 7.93$ &$ 87$ &$ -1.47$ &$ 7.33$ &$ 61$ &$  1.85$ &$ 7.79$  \\
\hline
\end{tabular}
\end{table*}

\section{Discussion}

The negative cosmic shear estimate that we find when considering all the fields, if not a statistical fluke, can only be
due to selection effects or systematics present in our data. We review in the following the possible 
causes of systematics and try to assess their impact.  

\subsection{Selection and weighting effects}

To verify the validity of the data reduction, which is somewhat different from the one used on the 
first 2 papers, we simulated 210 associations with an average of 24 galaxies per field using Skymaker (Erben et al. 2001) in a 
similar fashion
as described in Sect. 5.4 of HMS02. Each associations consists of 4 members with relative random shifts between 0 and 3 pixels 
which are coadded in the same way as the real data.  We found that, 
as for the archival data in HMS02, the final cosmic shear estimate is 2.4\% with a 3$\sigma$ significance as 
compared to the 2.6\% true shear introduced in the simulated galaxy catalogs. We estimate then that the reduction procedure 
is completely equivalent for our purposes and is not responsible for the negative shear estimate.

We investigated if this negative value could be the product of the way that fields
are selected. Since parallel imaging pointings are controlled by the observations
of the primary instrument, one could think that we may be biased towards a certain category of fields.
We did 100.000 random selections of 121 fields (the number of fields in HMS02) out of the 210 available 
and computed the mean of the cosmic shear estimator for these fields. The distribution of the obtained 
values is shown in Fig. 11. A value as large as the one obtained in the analysis of the archival data 
from HMS02 could be obtained in just 0.003\% of the realisations.
   
\begin{figure}
   \centering
   \includegraphics[width=8.5cm]{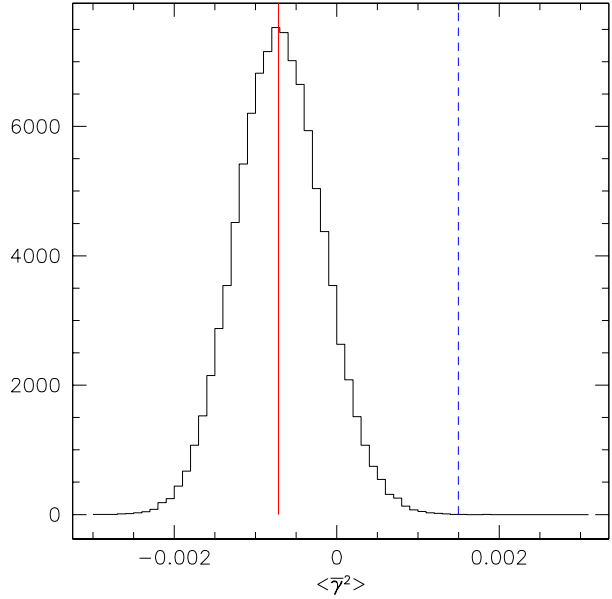}
 \caption{Distribution of the cosmic shear estimator calculted from 121 randomly selected fields out of the 210.
The vertical solid line represents the value of the estimator for this data, while the vertical dashed
line represents the value of the estimator for the data measured in HMS02}
\end{figure}

In Table 1, we summarize the results for different selections of fields, different cuts in 
$\Pg$, different weightings of individual galaxies and galaxy fields. 
If we calculate the estimator by varying the cuts in $\Pg$, no significant variation is obtained, which indicates
that the measurement is not dominated by a few noisy galaxies with large correction factors. 
The effect of the weighting of individual galaxies is more significant since the estimator becomes more negative
when a weighting is applied (with no difference between $w^\prime_\mathrm{NN}$ and $w_\mathrm{NN}$. 
If we vary the weighting of the galaxy fields, the dispersion is minimized for a Poisson noise weighting 
$W_\mathrm{f} = N$, though the estimator is less negative with $W_\mathrm{f} = N^2$. 
We observe, as it was the case in HMS02, that the shear estimate increases when we select fields with 
a larger number density of galaxies, which are typically deeper exposures for which we expect galaxies to 
be on average at a larger redshift and therefore the true shear signal to be higher. This effect was also 
seen in the archival data. It indicates then that even if our estimator is negative, a true shear signal 
may be present in our data.

\subsection{Hot pixels}

The impact of hot pixels can be seen in in the last 2 blocks of Table 1. When using objects with a larger 
r$_\mathrm{h}$, which are less affected by left-over hot pixels,  or fields where exposures are dithered by more than 
1 pixel (which can be cleaned of hot pixels), the cosmic shear estimate increases slightly. 
But since the number of galaxies is also reduced, the dispersion of the result is higher and therefore not significant.
We have to note also that for larger objects, the observed behaviour could be due also to CTE effects as discussed later. 

\subsection{PSF effects}

The effect of each individual star field PSF correction on the final shear estimate result is presented
on Table 2.
As seen in the first block of Table 2, even in the absence of PSF corrections, the cosmic shear estimate has a
negative value, and the full correction of the PSF just decreases even more the value of the estimate. 
We observe that the final result does not vary significantly between the different corrections and
that all results are well within the statistical error of the shear estimate. 
The effect of the PSF correction for most of the fields is rather small as can be seen in Fig. 12. 
As a test, we applied the PSF correction as estimated from the archival star fields 
to the Cycle 9 galaxy fields. And vice-versa, we applied PSF correction as estimated from the 
Cycle 9 star fields to the galaxy fields from HMS02. In both case we found no significant variation of the result
in the cosmic shear estimate. 

\begin{figure}
   \centering
   \includegraphics[width=8.5cm]{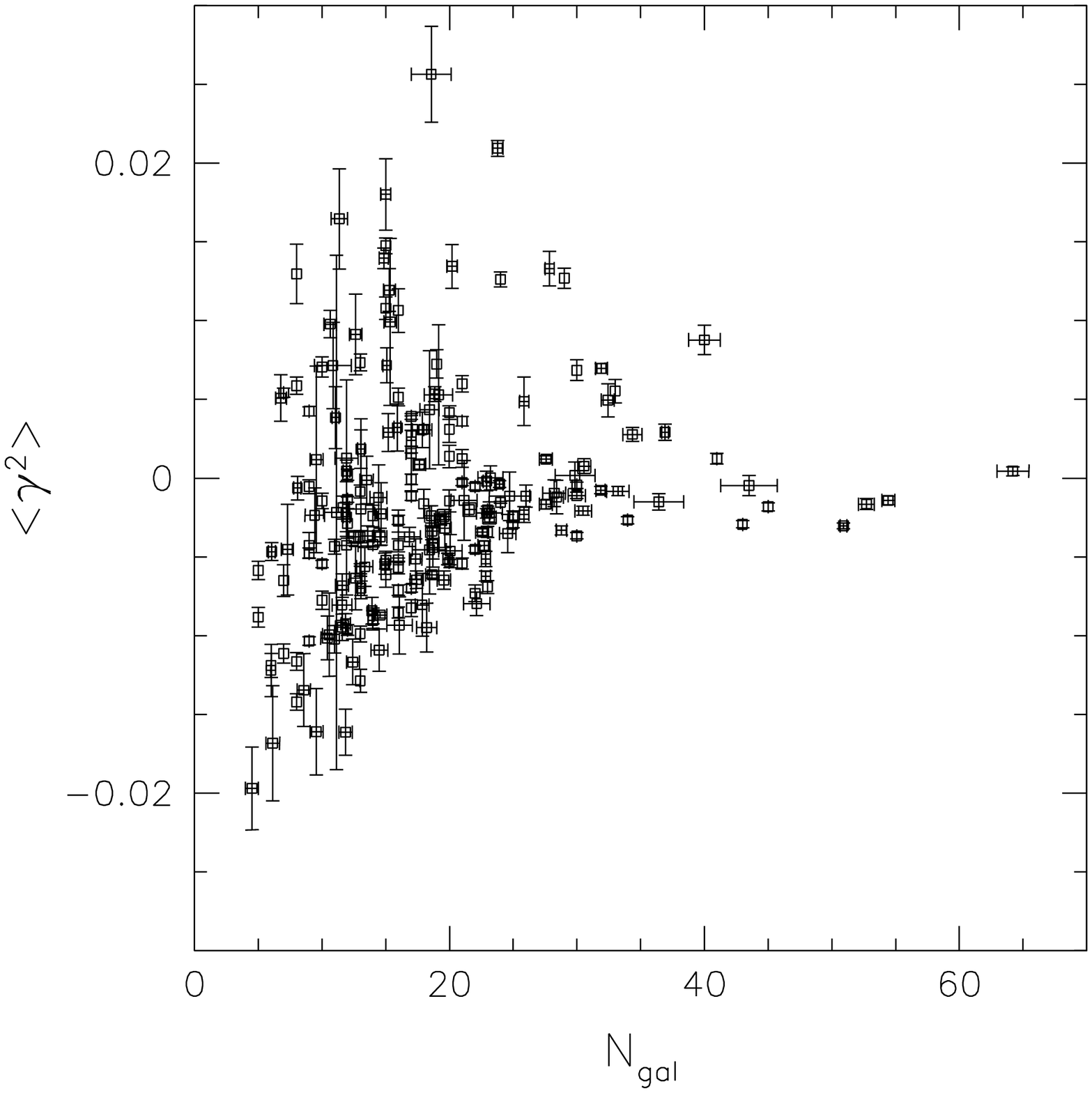}
 \caption{We show for each of the 210 galaxy fields the cosmic shear estimator as a function of the number
of galaxies. The vertical error bars indicate the 1$\sigma$ dispersion from the mean. The horizontal
error bars indicate the variation in the number of objects depending on the cut in $\Pg$}
\end{figure}

\begin{table*}[!t]
\caption{Results for the cosmic shear estimator for different PSF
corrections, weighting individual galaxies with $w=w^\prime_\mathrm{NN}$, requiring
$\Pg>0.2$ (or $P_\mathrm{sh}>0.2$), and weighting the galaxy fields by
$W_\mathrm{f}=N$. Note that even after the cut in $\Pg$ some galaxies
are left with unphysical ellipticities larger than one which were
excluded  from the analysis. This leads to the different number of
galaxy fields for $N\ge10$ and  $N\ge15$ in the
first block, where we show the results if we do not correct for PSF
effects:
the first column indicates if we use uncorrected (raw) ellipticities
($\e^\mathrm{raw}$) or anisotropy-corrected ellipticities
($\e^\mathrm{ani}$) and if we apply  the smearing correction ($\Pg$)
or not ($P_\mathrm{sh}$).  The first row gives the fully corrected
result (see Table 1) for reference.
The next block shows the results when we apply PSF corrections from
the individual star fields.}
\center
\begin{tabular}{|l|rrr|rrr|rrr|}
  \noalign{\smallskip}
  \noalign{\smallskip}
\hline
 \multicolumn{1}{|c|}{$e$, $P$} & \multicolumn{3}{c|}{all} &
  \multicolumn{3}{c|}{$N\ge10$} & \multicolumn{3}{c|}{$N\ge15$}\\
 \multicolumn{1}{|c|}{or}       & $N_\mathrm{f}$ & $\cs$ & $\sigcs$ &
  $N_\mathrm{f}$ & $\cs$ & $\sigcs$ & $N_\mathrm{f}$ & $\cs$ & $\sigcs$ \\
 \multicolumn{1}{|c|}{starfield} && $\times 10^{4}$ & $\times 10^{4}$
  && $\times 10^{4}$ & $\times 10^{4}$ && $\times 10^{4}$ & $\times 10^{4}$ \\
\hline
\hline
$\e^\mathrm{ani}$, $\Pg$           & $210$ &$ -7.16$ &$ 5.13$ &$184$ &$ -6.04$ &$ 4.97$ &$130$ &$  1.44$ &$ 5.36$ \\
$\e^\mathrm{raw}$, $\Pg$           & $210$ &$ -8.12$ &$ 5.07$ &$184$ &$ -7.08$ &$ 4.91$ &$131$ &$ -0.42$ &$ 5.30$ \\
$\e^\mathrm{ani}$, $P_\mathrm{sh}$ & $210$ &$ -1.25$ &$ 2.07$ &$203$ &$ -0.99$ &$ 2.08$ &$181$ &$ -0.06$ &$ 2.14$ \\
$\e^\mathrm{raw}$, $P_\mathrm{sh}$ & $210$ &$ -1.80$ &$ 1.99$ &$206$ &$ -1.68$ &$ 2.00$ &$186$ &$ -0.94$ &$ 2.05$ \\
\hline
o6969zaz0\_3\_ass & $210$ & $-7.59 $& $5.55$ & $182 $& $-6.22$ & $5.31 $& $127$ & $ 1.28 $& $5.83$ \\
o696nnmu0\_2\_ass & $210$ & $-5.24 $& $4.97$ & $186 $& $-4.33$ & $4.84 $& $133$ & $ 2.10 $& $5.26$ \\
o696surs0\_3\_ass & $210$ & $-7.39 $& $5.40$ & $182 $& $-6.02$ & $5.19 $& $128$ & $ 2.28 $& $5.68$ \\
o6fx9j010\_2\_ass & $210$ & $-5.90 $& $5.13$ & $186 $& $-4.94$ & $4.97 $& $132$ & $ 2.06 $& $5.44$ \\
o6fxc7f30\_4\_ass & $210$ & $-8.10 $& $5.52$ & $182 $& $-6.81$ & $5.32 $& $129$ & $ 0.73 $& $5.91$ \\
o6fxdeng0\_1\_ass & $210$ & $-8.95 $& $5.65$ & $182 $& $-7.72$ & $5.40 $& $125$ & $-0.90 $& $5.88$ \\
o6fxdmeo0\_2\_ass & $210$ & $-7.62 $& $5.43$ & $183 $& $-6.04$ & $5.22 $& $129$ & $ 1.25 $& $5.71$ \\
o6fxdsrr0\_2\_ass & $210$ & $-8.49 $& $5.48$ & $182 $& $-7.13$ & $5.29 $& $128$ & $ 0.88 $& $5.84$ \\
o6fxe5xq0\_3\_ass & $210$ & $-6.14 $& $5.07$ & $186 $& $-5.11$ & $4.92 $& $133$ & $ 1.16 $& $5.37$ \\
o6fxebok0\_3\_ass & $210$ & $-6.18 $& $5.11$ & $186 $& $-5.15$ & $4.98 $& $133$ & $ 1.24 $& $5.47$ \\
o6fxep010\_2\_ass & $210$ & $-6.84 $& $5.05$ & $186 $& $-5.81$ & $4.91 $& $133$ & $ 0.46 $& $5.37$ \\
o6fxfohi0\_2\_ass & $210$ & $-5.43 $& $4.81$ & $186 $& $-4.72$ & $4.69 $& $133$ & $ 1.58 $& $5.10$ \\
o6fxfuh50\_2\_ass & $210$ & $-5.98 $& $5.00$ & $186 $& $-5.05$ & $4.86 $& $132$ & $ 1.62 $& $5.30$ \\
o6fxheu90\_2\_ass & $210$ & $-5.64 $& $5.02$ & $186 $& $-4.73$ & $4.90 $& $133$ & $ 1.65 $& $5.33$ \\
\hline
\end{tabular}
\end{table*}

\subsection{CTE effects}

As stated in Sect. 2, the CTE of the STIS CCD has been degradating by about 15\% per year on average
since 1997 (Goudfrooij et al. 2002, Proffitt et al. 2002b). This degradation is characterized by a loss in the 
efficiency of the transfer of
charges in the Y direction, which is related to the distance of the pixel from the 
read-out amplifier, the number of charges of the pixel and the number of charges between 
the pixel and the read-out amplifier. This effect is responsible for a loss of flux for all objects, but
in particular for faint objects in a low background environment (Goudfroij et al. 2002). 
We expect then that, for our galaxies which are a few counts over the sky background, this may affect 
also their shapes by introducing a correlation between $e_1$ and the Y position of the galaxies. If we plot, like in Fig. 13, the
average $e_1$ for all the galaxies as a function of Y, we observe that the closer the galaxies 
are to the bottom of the chip, the more they tend to be aligned towards the Y direction. This correlation was not seen in
the archival data (PCE01). The fact that big galaxies, with $\rh > 5$, seem to be less affected than small galaxies by the 
systematics causing the estimator to be negative as shown in Table 1 is also consistent with the way CTE degradation acts.
 
To estimate the effect of CTE degradation on the cosmic shear estimate for our data, we simulated ten thousand times 
200 fields with 25 galaxies per field, with the observed ellipticity dispersion and with a mean 
$e_1= -0.01$. We added a y-depence to $e_1$ in order to simulate an average CTE degradation with the form: 
$e_1 = -0.15 + 0.3 \times y/2000$.
This is about 5 times the CTE degradation seen for the average of all the galaxies. 
For one set of simulations we had no cosmic shear effect and for a second set we added a shear of 2.5\%, and we 
computed the distribution of $\cs$ for each set. 
As can be seen in Fig. 14, the effect of CTE degradation is to lower $\cs$. However, this is a worst 
case scenario where we assumed that the mean corrected $e_1$ is smaller than the maximum CTE induced ellipticities.
Also, this average CTE degradation alone cannot produce a negative signal as large as the one observed even
in the case where the shear is 0. But since this effect depends not only on the position but also on the flux of each galaxy, 
the background as well as on the date when the images where taken, only a physical model of the CTE degradation which 
would take into account all those parameters could allow us to estimate its true impact (P. Bristow, private communication). 
We tried to correct the CTE degradation effect in the same way as in van Waerbeke et al. (2000), by adding a 
constant term to $e_1$ as a function of the Y-position (and/or background, surface brightness) of the galaxy in order to have the 
mean value of $e_1$ over all galaxies to be 0. This method does not work in this case and the value of cosmic shear 
estimator remains negative. To do a proper correction, it would be necessary to restore each single image 
using a physical model of the CTE degradation which is not available yet.

\begin{figure}
   \centering
   \includegraphics[width=8.5cm]{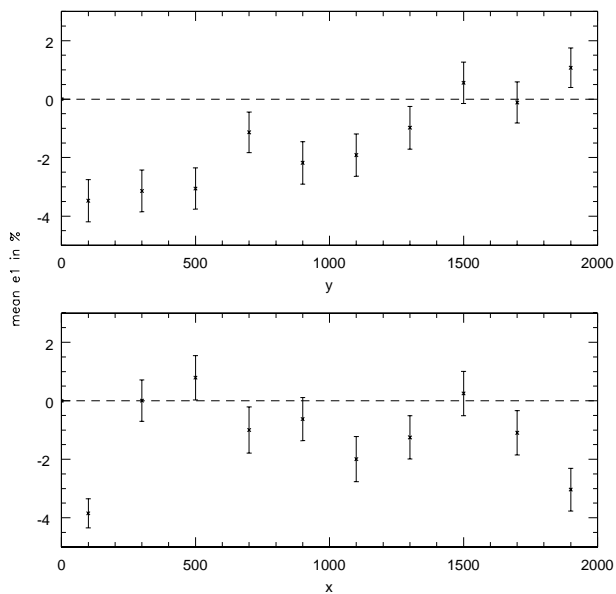}
 \caption{Average $e_1$ for all galaxies as a function of the X-position (bottom box) and  Y-position (top box)
in the field for bins of 200 subsampled STIS pixels. The error bars represent the variation on the mean.}
\end{figure}

\begin{figure}
   \centering
   \includegraphics[width=8.5cm]{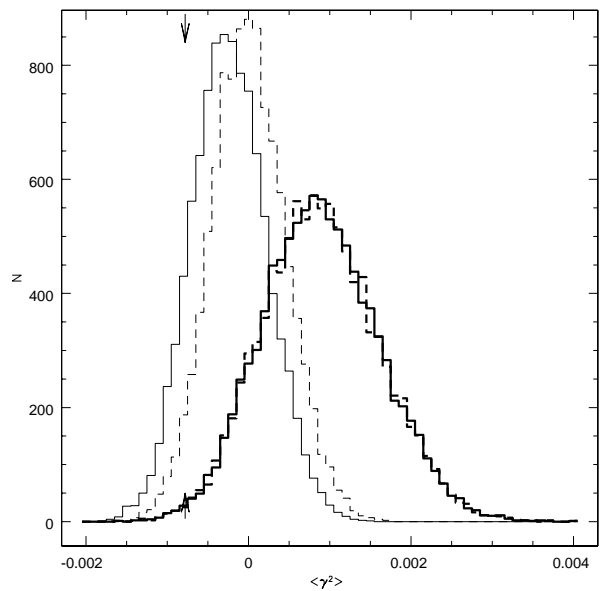}
 \caption{Cosmic shear estimator distribution for the simulated catalogs with a 2.5\% true shear (thick lines) and 
without (thin lines) shear. Full lines indicate the distribution for the catalogs including the average
CTE effect and the dashed lines the distribution for the catalogs without CTE effect. The arrow indicates the 
position of the measured cosmic shear estimate from the real data.}

\end{figure}

\section{Summary and Perspectives}

Following the analysis of the 210 galaxy fields selected from the Cycle 9 parallel proposal, we have found that
the cosmic shear estimate $\cs$ is consistently negative and therefore the cosmic shear signal is not detected on the scale
of 30$''$. Only fields with more than 15 galaxies per field yeld a positive $\cs$ though it is not statistically significant.
This is most likely due to the existance of several systematics that plague the STIS, however we cannot exclude that
this negative result is largely due to a statistical fluke.
We have studied several sources of noise and systematics that could contaminate this result: selection and weighting effects, hot pixels, 
PSF effects and degradation of the CTE. Neither of those seem to have by themselves the power to turn the estimate 
negative at the level observed in the presence of a significant signal (more than 1\%). But it appears also that the main effect 
for this negative result comes from the loss of charge transfer efficiency of the STIS CCD. 
Several groups (Goudfrooij \& Kimble 2002, Bristow et al.
2002) are trying to come up with physical models to correct for this effect and restore the images. Once these models 
will be publicly available, it will be interesting to try to model in detail the real effect of the 
charge transfer inefficiency on the shapes of individual galaxies.  
If it turns out that the difference observed between the results of data analysed in HMS02 and this one comes 
principally from the loss of CTE, then it would be important for future studies with 
space-based telescopes (SNAP, JWST) to prioritize these observations during the first periods of use of the instruments 
in order to minimize its impact.
  
Is it useful now to try to correct these STIS images for the cosmic shear measurements? Since the installation of 
the ACS camera onboard HST in March 2002, parallel data is also being acquired with it. The area covered by those observations 
already surpasses all the area covered with the analysed Cycle 9 pure parallel STIS data. We expect then the ACS data will 
better constrain the cosmic shear at the same scale than STIS since it will finally cover a much larger area at a similar or even 
greater depth without being contaminated, for the time being, by the effect of CTE loss. Also, since the ACS field of view 
is about 3.5$'$, it will be possible to cross-check its results with the available ground-based surveys at scales larger than the arcmin. 

\begin{acknowledgements}

This work was supported by German Ministry for Science and Education (BMBF) through the DLR under the project 50 OR106 and by the TMR
Network ``Gravitational Lensing: New Constraints on Cosmology and the Distribution of Dark Matter'' of the EC under contract
No. ERBFMRX-CT97-0172. RAEF is affiliated to the Space Telescopes Division, Research and Space Science Department, European Space Agency. 

\end{acknowledgements}


\begin{thebibliography}{}

\bibitem[Bartelmann \& Schneider]{BS01}
 Bartelmann, M. \& Schneider, P., 2001, Phys.Rep., 340, 291
\bibitem[Bertin \& Arnouts]{BA96}
 Bertin, E. \& Arnouts, S., 1996, A\&AS, 117, 393
\bibitem[Bristow et al. 2002]{bristow}
Bristow, P., Alexov, A., Kerber, F., Rosa, M., 2002, Proceedings of the 2002 HST Calibration Workshop, 
S. Arribas, A. Koekemoer, \& B. Whitmore, eds., 177
\bibitem[Crabtree et al. 1996]{crabtree} 
Crabtree, D.R., Durand, D., Gaudet, S., and Hill, N., 1996,
ADASS V, A.S.P. Conf. Ser., George H. Jacoby and Jeannette Barnes, eds., 101, 505. 
\bibitem[Erben et al.]{E01}
 Erben, T., van Waerbeke, L., Bertin, E., Mellier, Y. \& Schneider, P.,
 2001, A\&A, 366, 717
\bibitem[Fruchter \& Hook]{FH98}
 Fruchter, A.S. \& Hook, R.N., 2002, PASP, 114, 144
\bibitem[Gardner \& Satyapal]{gardner}
Gardner, J.P. \& Satyapal, S., 2000, AJ 119, 2589
\bibitem[Goudfrooij \& Kimble]{goudfrooij}
Goudfrooij, P. \& Kimble, R.A., 2002, Proceedings of the 2002 HST Calibration Workshop, S. Arribas, A. Koekemoer, \& B. Whitmore, eds., 105
\bibitem[H\"ammerle et al.]{HMS02}
 H\"ammerle, H., Miralles, J.-M., Schneider, P. et al., 2002, A\&A, 385, 743 (HMS02)
\bibitem[Hoekstra et al.]{Hea98}
 Hoekstra, H., Franx, M., Kuijken, K. \& Squires, G., 1998, ApJ, 504, 636
\bibitem[Hudson et al.]{Hu98}
 Hudson, M.J., Gwyn, S.D.J., Dahle, H. \& Kaiser, N., 1998, ApJ, 503, 531
\bibitem[Kaiser]{K92}
 Kaiser, N., 1992, ApJ, 388, 272
\bibitem[Kaiser]{K98}
 Kaiser, N., 1998, ApJ, 498, 26
\bibitem[KSB95]{KSB}
 Kaiser, N., Squires, G. \& Broadhurst, T., 1995, ApJ, 449, 460
\bibitem[Luppino \& Kaiser]{LK97}
 Luppino,~G.A. \& Kaiser,~N., 1997, ApJ, 475, 20
\bibitem[Micol et al. 1998]{micol} 
Micol, A., Pirenne, B., \& Bristow, P., 1998, ADASS VII, A.S.P. Conf., R. Albrecht, R.N. Hook \& H.A. 
Bushouse, eds., 145, 45
\bibitem[Paper~I]{PaperI}
 Pirzkal, N., Collodel, L., Erben, T., et al., 2001, A\&A, 375, 351 (PCE01)
\bibitem[Proffitt et al. 2002a]{proffitta}
Proffitt, C. R., Goudfrooij, P., Brown, T. M., et al., 2002a, Proceedings of the 2002 HST Calibration Workshop, S. Arribas, A. Koekemoer, \& B. Whitmore, eds., 97
\bibitem[Proffitt et al. 2002b]{proffitb}
Proffitt, C., et al. 2002, "STIS Instrument Handbook", Version 6.0, Baltimore: STScI
\bibitem[Rhodes et al.]{Rea01}
 Rhodes, J., Refregier, A. \& Groth, E.J., 2001, ApJL, 552, 85
\bibitem[van Waerbeke et al.]{vWea00}
 van Waerbeke, L.,  Mellier, Y., Erben, T., et al.,
 2000, A\&A, 358, 30


\end{thebibliography}
\end{document}